\documentclass[twocolumn,showpacs]{revtex4}

\usepackage{graphicx}
\usepackage{dcolumn}
\usepackage{amsmath}

\makeatletter
\def\btt#1{\texttt{\@backslashchar#1}}%
\DeclareRobustCommand\bblash{\btt{\@backslashchar}}%
\makeatother

\begin{document}


\title[Timelapse]{TIMELAPSE}

\author{Stefano De Leo}
\email{deleo@ime.unicamp.br}
\affiliation{Department of Applied Mathematics, University of Campinas\\ 
PO Box 6065, SP 13083-970, Campinas, Brazil\\
}

\author{Pietro Rotelli}
\email{rotelli@le.infn.it}
\affiliation{Department of Physics and INFN, University of Lecce\\ 
PO Box 193, I 73100, Lecce, Italy}

\date{November 29, 2001}

\begin{abstract}
\vspace*{0.5cm}
We discuss the existence in an arbitrary frame of a finite time for the 
transformation of an initial quantum state into another e.g. in a decay. 
This leads to the introduction of a timelapse $\tilde{\tau}$ in analogy with 
the lifetime of a particle. An argument based upon the Heisenberg uncertainty 
principle suggests the value of $\tilde{\tau}=1 / M_0$. Consequences for the 
exponential decay formula and the modifications that $\tilde{\tau}$ introduces
into the Breit-Wigner mass formula are described.
\end{abstract}

\maketitle




The subject of this paper concerns the exponential decay law and the 
Breit-Wigner (BW) mass distribution formula. These formulas are a standard 
part of particle physics and can indeed be connected by a simple transform, 
as is taught in many undergraduate physics courses~\cite{HM}. 
In recent years the 
validity of the B-W has been demonstrated to an unprecedent degree by the 
LEP data 
and analysis upon the Z gauge particle~\cite{DRE}. 
After substantial theoretical 
corrections for radiative effects the predicted theoretical widht 
(assuming three light neutrinos) and the experimental value, 
{\em based upon a B-W fit}, agree to better than one per cent. We will be 
interested later in the small discrpancy but our first observation is that 
the agreement is quit impressive. 
These facts are somewhat surprising because neither the exponential 
decay law nor the B-W mass curve is predicted rigorously within 
Quantum Mechanics (QM). On the contrary we have precise QM objections to 
the former~\cite{KHA,SW} and only approximate derivations of the 
latter~\cite{WW,COH}.
Nor does a field theoretical treatment change 
the situation. To some these QM results relegate our two formulas to 
little more than phenomenological games. We in the other hand start from 
these two formulas and argue for a modification which will in part 
reconcile the decay law with QM, and provide an explanation for the 
discrpancy in the Z widht described above.

One of the implicit assumptions in particle physics is that decays occur
instantaneously. However, as Einstein has taught us, istantaneity, 
for anything other than a point, can 
at best be valid in a single
Lorentz frame. Thus even if, say in its rest frame, a particle decayed 
instantaneously, a general observer would find different times for decays at 
different points within the wave packet. Of course, this could only be 
determined in a statistical sense since the wave function is not an 
observable. Thus, in general, there will
exist times during which the quantum state is neither the initial nor the 
final state but a {\it linear combination} of both which tends towards the 
later with increasing time. We shall call a measure of this time interval
the ``timelapse''.

There is only one exception to the above 
observations, a measurement process may involve (ideally) the localization 
in space and time of a particle. This collapse of the wave function is
instantaneous for all observers (it defines the corresponding event in
each frame) and is fundamentally irreversible since the creation of a
particle at a given space time point cannot instantaneously inflate to
finite space regions without violating the limiting velocity of
light. The collapse of the wave function is a subject of great
interest in itself but will not concern us further in this paper.

Why cannot we avoid the discussion of 
timelapse by considering particle or state creation to be a delta
function in space and time? Firstly because in many practical problems
we know this not to be the case, such as for a particle {\it trapped}
within a potential well e.\,g. a muon within a muonic atomic state.
Secondly because we often know or desire to
study particles which approximate energy-momentum eigenstates and this
implies {\it large} spatial dimensions.

An earlier introduction of a type of timelapse is contained in the book
of Jackson~\cite{JAC}. In one of the classical derivations of essentially 
quantum effects, Jackson
introduces the ``formation time'' of the electron in nuclear beta
decay. Arguing that the outgoing energetic electron could be
considered to have been accelerated from rest to its final velocity
over a finite formation time, he calculates the induced
spectrum of radiation by the electron, the {\it inner
bremsstrahlung}. Jackson also notes that the same effect would result if
the charge were {\it created} over the same time interval. Invoking the
uncertainty principle, he evaluates this time interval $\Delta t$
as
\begin{equation}
\label{dtde}
\Delta t \sim 1 / E~,
\end{equation}
where $E$ is the electron's energy. Now, while acknowledging precedence
for the idea of a formation time to Jackson, his approach is
significantly different from ours. The use of the particles energy in 
Eq.~(\ref{dtde}) implies that different particles take different times
for acceleration. Of course, 
the antineutrino does not contribute to the inner bremsstrahlung and  
the heavy nucleon contributions are negligible, so 
this may appear an academic question.
However, to us, it is obvious that the same timelapse must occur for
each of the particles in the final state, independent of their final
energy or momentum. It is not conceivable that the outgoing electron has
been created with certainty close to one while the antineutrino is
perhaps to all extents still to be created. 
We shall take
care to define for {\it each decay} a common timelapse that depends
at most upon the kinematics of the initial system  
in a preferential Lorentz frame. Of course, one must
also guarantee  
that the timelapse of the outgoing state coincides with that of 
the vanishing incoming state.

Is the timelapse a function of the 
spatial localization of the quantum state? 
We believe not. There {\it is} a $\Delta t$ directly connected to $\Delta x$
but this is what we might call ``passage time''. Consider a single
particle with a sufficiently large $\Delta x$ to be an approximate 
$(E,p,0,0)$ eigenstate (for simplicity we take its momentum to be
along the $x$ axis). Using the uncertainty principle 
$\Delta x \Delta p \sim 1$ and the Einstein relation $E^2 = p^2 + M^2$,
we find
\[
\Delta x \rightarrow \Delta p \rightarrow  \Delta E =  p \Delta p / E \sim 
v / \Delta x 
\]
hence
\begin{equation}
\Delta t \sim \Delta x /v~,
\end{equation}
where $v$ is the velocity of the particle. Thus $\Delta t$ is for our 
wave packet a measure of the time it takes to pass a given $y$-$z$
plane, hence the name passage time.
This $\Delta t$ obviously has nothing to do with a timelapse, and
indeed becomes infinite in the particle rest frame. On the other hand,
we expect, from the approximate validity of the exponential decay law, 
that timelapse
must be small compared to the lifetime of a particle  in
any frame, see below.

We  believe that timelapse must be a close relative to lifetime. As for
lifetime, we will define it in the rest frame of an initial single
particle state, or more precisely the frame in which the average
velocity is null. 
In analogy with the lifetime $\tau$, 
we shall denote the timelapse for a process by $\tilde{\tau}$. 
What does the Heisenberg  uncertainty principle
tell us? Well, we have excluded a connection to $\Delta E$ and we can
also exclude the half width $\Delta M$ for a decay particle since this is
reserved for $\tau$
\begin{equation}
\label{dm}
\tau = 1 / \Delta M~.
\end{equation}
This leaves us with essentially only one choice
\begin{equation}
\label{m0}
\tilde{\tau} = 1 / M_0
\end{equation}
for a decaying particle with central mass $M_0$. 

Now a decay of a composite particle such as a $J/\psi$ may be considered an
annihilation and/or interaction at the quark level. Thus, timelapses for 
interactions should also be defined. The natural choice for $\tilde{\tau}$
in these cases is
\begin{equation}
\label{ecm}
\tilde{\tau} = 1 / E_{CM}~,
\end{equation} 
where $E_{CM}$ is the center of mass energy. 
Equation (\ref{ecm}) would automatically include Eq.~(\ref{m0})
if it where not for the fact that a given decay may occur at a mass
diverse from $M_0$ due to the existence of mass curves. To reconcile
the two, we should modify Eq.~(\ref{m0}) to read
\begin{equation}
\label{m}
\tilde{\tau} = 1 / M~,
\end{equation}
but in the subsequent applications, in this paper
we will employ Eq.~(\ref{m0}) for simplicity.

How does the existence of a $\tilde{\tau}$ modify the
exponential decay law? This can easily
be derived after assuming a given analytic form for a state during
timelapse. For simplicity and in analogy with the original decay law,
we shall assume this to be an exponential form. Exponential decreasing
$\exp [- t / \tilde{\tau}]$ for the incoming state and its complement, 
$1 - \exp [- t / \tilde{\tau}]$, for the corresponding outgoing
state.
Thus when we consider an ensemble of $N(t)$ particles with a given
lifetime $\tau$ and timelapse $\tilde{\tau}$ we must divide them into
two classes: $N_u$, the number of undecayed particles, and $N_d$,
those that have begun the decay process but have still a residual probability
of being found in a measurement of $N(t)$. $N_d$ would be zero if
instantaneous decay  ($\tilde{\tau}=0$) were valid.

The differential equation that governs $N_u (t)$ is the standard one
\begin{equation}
\label{nu}
\mbox{d} {N}_{u}(t) = - \, \frac{1}{\tau} \, N_{u}(t) \, \mbox{d}t
\end{equation}
while that for $N_d (t)$ must allow for a source term proportional to
d$N_u (t)$,
\begin{equation}
\label{nd}
\mbox{d} {N}_{d}(t) = - \, \frac{1}{\tilde{\tau}} \, N_{d}(t) \, \mbox{d} t 
- \mbox{d}{N}_{u}(t)
\end{equation}
the negative sign in front of the last term is the correct one since
$\mbox{d}{N}_u (t) < 0$ for $\mbox{d}t > 0$.

Now solving these coupled equations and using as initial condition 
$N_{d}(0)=0$, we find
\begin{equation}
\label{mdl}
N(t) = \frac{N(0)}{ \tau - \tilde{\tau}} \, \, \left[ \, \tau 
\, \exp \left( - \, \frac{t}{\tau} \, \right) - \tilde{\tau}  
\, \exp \left( - \, \frac{t}{\tilde{\tau}} \, \right) \right] ~. 
\end{equation}
This is the modification of the standard exponential decay law that
our timelapse $\tilde{\tau}$ introduces. Note that for
$\tilde{\tau}=0$
we obtain the single exponential form and 
the limit $\tau = 0$ yields an exponential decay with
lifetime $\tilde{\tau}$. Whence $\tilde{\tau}=1/M_0$ sets a lower limit to 
the effective lifetime and hence an upper limit to the half width.

A simple calculation yields for the effective lifetime $\tau_{eff}$
\begin{equation}
\label{elfa}
\tau_{eff} = \left( \tau^{3} - \tilde{\tau}^{3} \right) / 
\left( \tau^{2} - \tilde{\tau}^{2} \right)~.
\end{equation}
For $\tilde{\tau} \ll \tau$, we can write
\begin{equation}
\label{elf}
\tau_{eff} = \tau \, \left[ 1 + \epsilon^2 + 
\mbox{O} \left( \epsilon^3 \right) \right]~,
\end{equation}
where $\epsilon=\tilde{\tau}/ \tau$.  
However, we note that the decay rate $\Gamma$ remains
connected to $\tau$, $\Gamma = 1 / \tau$. This follows from our
assumption that $\tilde{\tau} \sim 1 / M_0$ and hence is {\it
independent} of any interaction coupling constants in contrast to 
$\Gamma$ and $\tau$ which obviously are directly dependent.

Another, very interesting, observation is that, for very small $t$, 
Eq.~(\ref{mdl}) has no linear term in $t$. Indeed for $t \ll
\tilde{\tau} \, , \, \tau$
\begin{equation}
\label{mdl2}
 N(t) =  N(0) \left[ 1 - \frac{\, t^2}{2 \tau \tilde{\tau}} + 
\mbox{O}\left(t^3 \right) \right]~.
\end{equation}
This reconciles, for short times, our modified decay law (no longer a single
exponential) with basic quantum mechanical arguments~\cite{NAM,ARN,TQM} which 
have led,
amongst other things, to the so called quantum Zeno effect~\cite{ZEN,ZEN2}.
This at least in principle allows
$\tilde{\tau}$ to be calculated, from Eq.~(\ref{mdl2}) and the quantum
mechanical result $P(t) = 1 - t^2 ( \Delta H )^2 + ...$~\cite{TQM2},
specifically $\tilde{\tau} = 1 / \left[ 2 \tau ( \Delta H )^2 \right] = 
\Gamma / \left[ 2 ( \Delta H )^2 \right]$.


\begin{figure}
\includegraphics[width=8.5cm, height=10cm, angle=0]{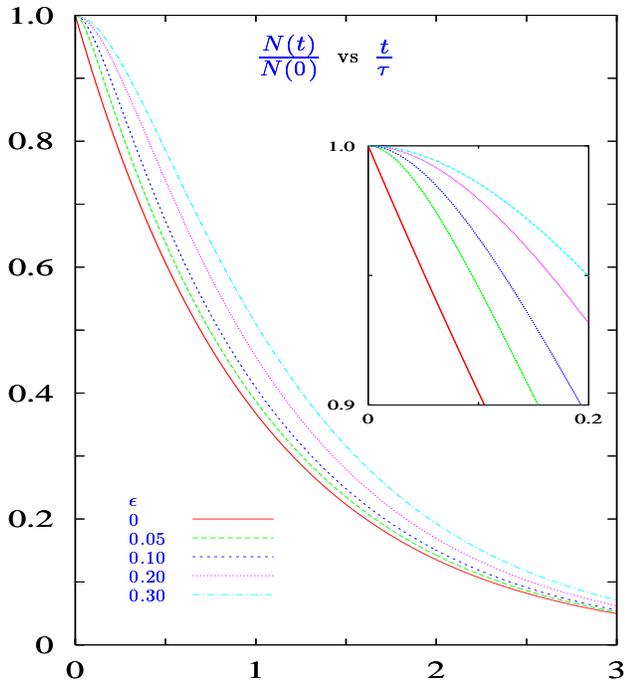}
\caption{The decay law $N(t)/N(0)$ versus $t/\tau$ for various values
of $\epsilon$.}
\label{fig_1}
\end{figure}


In Fig.~(\ref{fig_1}), we show the modification of 
$| \psi(t)|^2\equiv P(t)$ for various $\epsilon$  
values. In this 
plot the increase of the effective lifetime is evident, as is the
annulment of $\mbox{d}{P}(t) / \mbox{d}t$ (insert) 
for $t \rightarrow 0$, source of 
the quantum Zeno effect.
The direct measurement of 
$| \psi(t) |^2$ is often 
possible (e.g. in muon decay). However, for muons 
$\epsilon \sim 3 \times 10^{-16}$ so that no
effect due to $\tilde{\tau}$ could ever be detected.

Let us now calculate the modification in the standard Breit-Wigner 
mass formula produced by Eq.~(\ref{mdl}). We have
\[
| \psi(t) |  \propto  \exp \left( - \, \frac{t}{2 \tau} \, \right) \, 
\left\{ 1 - \epsilon \, \exp \left[ \, \frac{t \left( \epsilon - 
1 \right)}{  \epsilon \tau} \, \right] \right\}^{1/2}~.
\]
Hence,
\begin{eqnarray*}
\chi(x) & = & \int_{0}^{\infty} \, \mbox{d}t \, \exp 
\left( i  \, \frac{x}{2 \tau} \, t   
\right) \, | \psi(t) |\\
& \propto  & \frac{1}{x + i} - \sum_{n=1}^{\infty} \, \frac{\, n!!}{2^n \, n!}
\, \frac{\epsilon^n}{x + i \, a_n}
\end{eqnarray*}
where $x=2 \tau (M-M_0)$ and $a_n=1+ 2n ( 1-\epsilon) / \epsilon$. 
Treating $\epsilon$ as a small quantity, we may perform an 
analytic calculation of $\left| \chi(x) \right|^2$
to lowest order in $\epsilon$. We find for our modified Breit-Wigner ($MBW$)
\begin{widetext}
\begin{equation}
\label{mbw}
MBW \equiv  \left| \chi(x) \right|^2  
        =  BW \, \left\{ 1   
- \epsilon \, \, \frac{x^2+a_1}{x^2+a_1^2} + \frac{\, \epsilon^2}{4} \, \left[
\, \frac{7}{2} + 
\frac{x^2+1}{x^2+a_1^2} - \frac{x^2+a_2}{x^2+a_2^2} \right] + \mbox{O} \left(
\epsilon^3 \right) 
\right\} 
\end{equation} 
\end{widetext}
where 
$BW = 2 \tau M_0 / \left\{ 
\left[ \pi /2 + \arctan \left( 2 \tau M_0 \right) \right] 
\left(x^2 + 1 \right) \right\}$ is the standard Breit-Wigner.

Now $\epsilon$ is so small for almost all weak or electromagnetic decays that
one might think to pass directly to the strong decays in the search of 
evidence for a $MBW$. However, 
of all the weak processes a special role is played by the decays of the  heavy
intermediate vector bosons $W$ and $Z$. These have widths of {\it several GeV}
and thus correspond to 
$\epsilon$ of a few $\%$. Furthermore, the data upon 
the $Z$ is particularly precise with errors in $M_{Z}$ and $\Gamma_Z$ of order
$10^{-5}$. The LEP data have yielded such precise results that there is even
a two standard deviation from theory in $\Gamma_Z$. This is expressed
by two equivalent numbers~\cite{DRE}
\[
\Gamma_{inv} = - 2.7^{\, +1.7}_{\, -1.5} ~ \mbox{MeV}
\]
and/or 
\[ 
N_\nu = 2.9841 \pm 0.0083
\]
Now for the $Z$ we can apply the small $\epsilon$ formula given above since 
$\epsilon_Z=2.7 \%$.
From this formula (see also Fig.~\ref{fig_2} below) we readily see that:\\ 
(1) The maximum modification to the {\em underlying}
Breit-Wigner ($\tilde{\tau}=0$)
is at $M=M_{0}$ i.e. at the peak. This effect is an increase of the peak
value by $1 + 3 \epsilon^2 / 8$. Note, however, that this underlying
Breit-Wigner must not be 
confused with the {\em best fit} Breit-Wigner need 
to the data in the presence of a 
non negligible $\tilde{\tau}$.  
\\
(2) The halfwidth of the underlying Breit-Wigner and the modified 
Breit-Wigner are almost the same for small $\epsilon$. That of the $MBW$ 
is reduced by a factor of order 
$\epsilon^3$ and not $\epsilon^2$ as might have been expected from 
Eq.(\ref{elf}) for the lifetime modification.

The smallest errors of the curve are around the peak value $M=M_0$.
In a fit with a Breit-Wigner to data in accord with our modified curve one 
would be inclined to raise the peak value with consequently the same 
decrease in percentage of the halfwidth. Hence, we expect the {\em fitted 
Breit-Wigner} to yield a width lower than ours by, at most,  the factor 
$1 - 3 \, \epsilon^2 / 8$. Hence our estimate of $\Gamma_{inv}$ is
\[  - 0.7 \, \mbox{MeV} \le \Gamma_{inv} < 0 \]
less than $1/4$ of the measured central value. It is amusing to note that 
using the 
result $\Gamma_{eff}=(1+ \epsilon^2) \,  \Gamma$ one might have expected an 
effect on 
$\Gamma_{inv}$ in good agreement with 
the experimental value.


\begin{figure}[bp]
\includegraphics[width=8.8cm, height=10cm, angle=0]{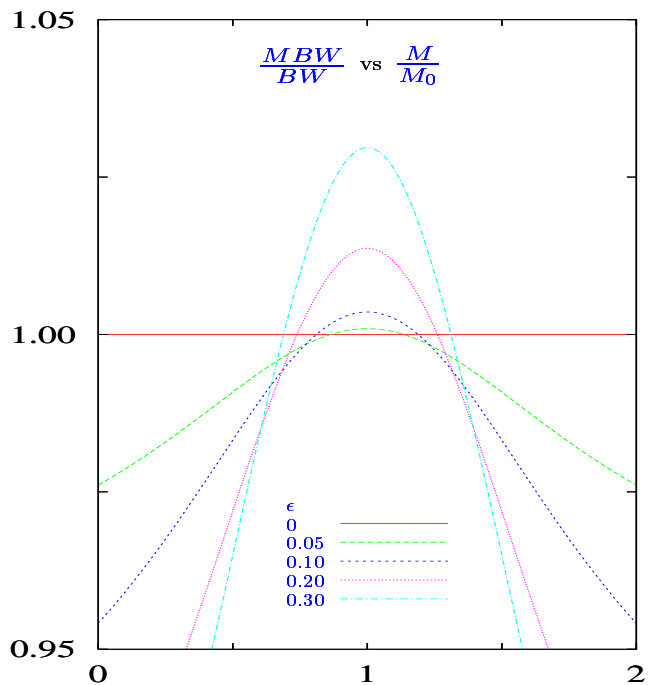}
\caption{Numerical calculation of the ratio of the modified and 
standard Breit-Wigner mass formulas 
$MBW/BW$ versus $M/M_0$ for different values of $\epsilon$.}
\label{fig_2}
\end{figure}


The modified Breit-Wigner can also be calculated numerically for any
$\epsilon$ and in Fig.~(\ref{fig_2}) we show the ratio of this with
the underlying Breit-Wigner for various $\epsilon$ values. The width is
indeed reduced.
From Fig.~(\ref{fig_2}), we see that, for $\epsilon \leq 0.3$ and such that
$ 1.4 > M/M_0 > 0.6$, the main modification indeed occurs at $M=M_0$ and is an
increase of the order of a few \%  or less. This means that no significant
evidence for the existence of $\tilde{\tau}$ from mass curves is possible
until the individual errors of the data points are of this order or better.

Obviously in looking for evidence for our modified Breit-Wigner we
are led to consider the 
largest $\epsilon$ values available. This means particles
with strong interaction decays. For example the $\rho(770)$ where 
$\epsilon  \sim 20 \%$.
However, the best data points for the  $\rho$~\cite{CAP}  
are somewhat dated and are not 
yet precise enough to yield evidence for a $\tilde{\tau}$. 

It is natural to extend the concept of timelapse, from the realm of decays to
interactions in general. We have already anticipated that in theses cases
$\tilde{\tau}=1/E_{CM}$. However what is $\tilde{\tau}$ to be compared with. 
What plays here the role of $\tau$? The only thing available is 
$\sqrt{\sigma}$ the square root of the cross-section. For numerical 
comparisons, we recall that $\sqrt{60 \, \mbox{mb}} \sim 10^{-23} \, \mbox{s}$
while $1 \, \mbox{Gev} \sim 6.6 \, 10^{-25} \, \mbox{s}$.

Finally, we wish to discuss briefly our stimulus for this investigation,
which seems at first sight far removed from the content of this
paper. In {\em oscillation studies} some authors insist that a single time
interval is involved. The argument is essentially that both the
creation of say a flavor neutrino and its detection, possible as a
different flavor, occur at fixed times. Now, as we explained in our
introduction, such a situation, {\it instantaneous creation}, could at most
be valid in a unique Lorentz frame which improbably coincides with
the laboratory. However, the existence of an
intrinsic timelapse $\tilde{\tau}$ would imply that instantaneous creation
is a myth {\em in any frame}. In practice this means that in interference
studies we must deal with multiple times in a similar way that the
{\it slippage} of interfering wave packets obliges us to consider
multiple distance intervals between creation and observation.

In conclusion, we have argued that in an arbitrary frame a wave packet
will take a finite time to ``grow'' to, or decay from, its full normalized 
value. This
encouraged us to postulate the existence in the preferential center of
mass frame of an intrinsic timelapse $\tilde{\tau}$ in analogy with
$\tau$. Such an assumption leads to a modification of the decay
formula and consequently of the Breit-Wigner mass formula. We have also
suggested that $\tilde{\tau}=1 / E_{CM}$ on the basis of the Heisenberg
uncertainty principle. In practice the modification of the decay
formula is not experimentally detectable. However, it has, as an aside, 
reconciled for small times the decay 
law with basic quantum mechanical arguments ( at
least within the hypothesis of an exponential dependence upon
$\tilde{\tau}$). The Breit-Wigner mass formula is a more practical
tool for detecting a $\tilde{\tau}$. 
Comparing the fits to the data upon $\rho$ decay suggest that 
with improved experiments (precision of the order of $10^{-3}$) we could
distinguish between the standard and modified Breit-Wigner. Note that
with $\epsilon = \Delta M / M_0$ the modified version has no extra free
parameters. We may simply compare the best $\chi^2$ fits of both
to the data. At the moment the most promising source for evidence of a 
$\tilde{\tau}$ appears to be in the $Z$ decay. Timelapse provides a 
justification for the existence of a negative $\Gamma_{inv}$. But
we must remember that this is experimentally only a two sigma effect.

We thank R.~Anni and G.~Co' for interesting discussions. One of the
authors (SdL) is grateful for the kind hospitality at the Physics 
Department of Lecce University, where the paper was written. This work
was partially supported by the FAEP (UniCAMP).



\begin{thebibliography}{99}

\bibitem{HM}
F. Halzen and A. D. Martin,
{\it quarks and Leptons: An Introductory Course in Particle Physics}
(John Wiley \& Sons, New York, 1984).

\bibitem{DRE}
J. Drees, 
{\it Review of Final LEP Results or A Tribute to LEP}, hep-ex/0110077. 

\bibitem{KHA}
L. A. Khalfin,
Phys. Lett. B {\bf 112}, 223 (1982).

\bibitem{SW}
Y. N. Srivastava and A. Widom,
Lett. Nuovo Cimento {\bf 37}, 267 (1983).


\bibitem{WW}
V. F. Weisskopf and E. Wigner,
Z. Physik {\bf 63}, 54 (1930).


\bibitem{COH}
C. Cohen-Tannoudji, B. Diu and F. Lalo\''e,
{\it Quantum Mechanics}, 
(John Wiley \& Sons, New York, 1977), Chap. 13.


\bibitem{JAC}
J. D. Jackson, 
{\it Classical Electrodynamics} 
(John Wiley \& Sons, New York, 1975), Chap. 15.

\bibitem{NAM}
M. Namiki and N. Mugibayashi,
Prog. Theor. Phys. {\bf 10}, 474 (1953). 

\bibitem{ARN}
E. Arnous and S. Zienau, 
Helv. Phys. Acta {\bf 34}, 279 (1951). 

\bibitem{TQM}
H. Nakazato, M. Namiki and S. Pascazio,
Int. J. Mod. Phys. B {\bf 10}, 247 (1996). 

\bibitem{ZEN}
B. Misra and E. C. G. Sudarshan, 
J. Math. Phys. {\bf 18}, 756 (1977). 


\bibitem{ZEN2}
P. Facchi, H. Nakazato and S. Pascazio,
Phys. Rev. Lett. {\bf 86}, 2699 (2001). 

\bibitem{TQM2}
P. Facchi and S. Pascazio,
Physica A {\bf 271}, 133 (1999). 


\bibitem{CAP}
L. Capraro {\it et al.},
Nucl. Phys.  {\bf B288}, 659 (1987). 



\end{thebibliography}

\end{document}